\documentclass[a4paper,prl,showpacs,twocolumn,superscriptaddress]{revtex4-1}
\usepackage{graphicx,hyphenat,braket}
\usepackage{color}
\usepackage{amsmath}

\renewcommand{\vec}[1]{\mathbf{#1}}
\providecommand{\operatorname}[1]{\mathop{\mathrm{#1}}\nolimits}
\providecommand{\e}{\operatorname{e}\hspace{-0.15em}}
\newcommand{\ii}{{\rm i}} 

\usepackage{xr}

\begin{document}

\title{Nonreciprocal magnons and symmetry-breaking in the noncentrosymmetric antiferromagnet}

\author{G.~Gitgeatpong}
\affiliation{Department of Physics, Faculty of Science, Mahidol University, Bangkok 10400, Thailand}
\affiliation{ThEP, Commission of Higher Education, Bangkok, 10400, Thailand}
\affiliation{Physics Program, Faculty of Science and Technology, Phranakhon Rajabhat University, Bangkok 10220, Thailand}

\author{Y.~Zhao}
\affiliation{Department of Materials Science and Engineering, University of Maryland, College Park, Maryland 20742, USA}
\affiliation{NIST Center for Neutron Research, National Institute of Standards and Technology, Gaithersburg, Maryland 20899, USA}

\author{P.~Piyawongwatthana}
\affiliation{Department of Physics, Faculty of Science, Mahidol University, Bangkok 10400, Thailand}

\author{Y.~Qiu}
\affiliation{NIST Center for Neutron Research, National Institute of Standards and Technology, Gaithersburg, Maryland 20899, USA}

\author{L.~W.~Harriger}
\affiliation{NIST Center for Neutron Research, National Institute of Standards and Technology, Gaithersburg, Maryland 20899, USA}

\author{N.~P.~Butch}
\affiliation{NIST Center for Neutron Research, National Institute of Standards and Technology, Gaithersburg, Maryland 20899, USA}

\author{T.~J.~Sato}
\affiliation{IMRAM, Tohoku University, Sendai, Miyagi 980-8577, Japan}

\author{K.~Matan}
\email[]{kittiwit.mat@mahidol.ac.th}
\affiliation{Department of Physics, Faculty of Science, Mahidol University, Bangkok 10400, Thailand}
\affiliation{ThEP, Commission of Higher Education, Bangkok, 10400, Thailand}

\date{\today}

\begin{abstract}
Magnons, the spin-wave quanta, are disturbances that embody a wave propagating through a background medium formed by ordered magnetic moments. In an isotropic Heisenberg system, these disturbances vary in a continuous manner around an ordered spin structure, thus requiring infinitesimal energy as a wavevector approaches a magnetic zone centre. However, competing anisotropic interactions arising from broken symmetry can favour a distinct static and dynamic spin state causing a shift of the minimum point of the magnon dispersion to a nonreciprocal wavevector~\cite{motome}. Here we report the first direct evidence of these nonreciprocal magnons in an antiferromagnet.  In the antiferromagnet we investigated, namely, noncentrosymmetric $\alpha$-Cu$_2$V$_2$O$_7$, they are caused by the incompatibility between anisotropic exchange and antisymmetric Dzyaloshinskii-Moriya interactions resulting in competing collinear and helical spin structures, respectively. The nonreciprocity introduces the difference in the phase velocity of the counter-rotating modes, causing the opposite spontaneous magnonic Faraday rotation of the left- and right-propagating spin-waves.  The breaking of spatial inversion and time reversal symmetry is revealed as a magnetic-field-induced asymmetric energy shift, which provides a test for the detailed balance relation.
\end{abstract}


\maketitle

While symmetry plays a central role in imposing uniformity on the fundamental laws of nature~\cite{feynman,noether}, symmetry breaking introduces ``the texture of the world''\cite{Gross} by adding layers of complexity to the physical laws. In condensed matter systems, symmetry and a lack of it determine the underlying interactions of the governing Hamiltonian. In particular, the absence of spatial inversion symmetry in magnetic systems causes the relativistic spin-orbit coupling, which gives rise to many intriguing phenomena such as the spin Hall effect~\cite{Sinova}, topological insulators~\cite{Qi}, multiferroics~\cite{Fiebig}, and noncentrosymmetric superconductors~\cite{Bauer:2004da}, to acquire antisymmetric Dzyaloshinskii-Moriya (DM) interactions~\cite{Dzyaloshinsky,Moriya}. For noncentrosymmetric $\alpha$-Cu$_2$V$_2$O$_7$, the crystal structure breaks spatial inversion symmetry~\cite{Calvo1975,Robinson1987}, and the antiferromagnetic ordering below $T_N=33.4$~K~\cite{ganatee,Lee:2016} breaks time reversal symmetry. The simultaneous breaking of both symmetries sets the stage for the intertwining electric and magnetic properties~\cite{Sannigrahi,Lee:2016} and for the existence of toroidal moments~\cite{Hayami2,Hayami3,Ederer}.

\begin{figure*}
\begin{center}
\includegraphics[height=10cm]{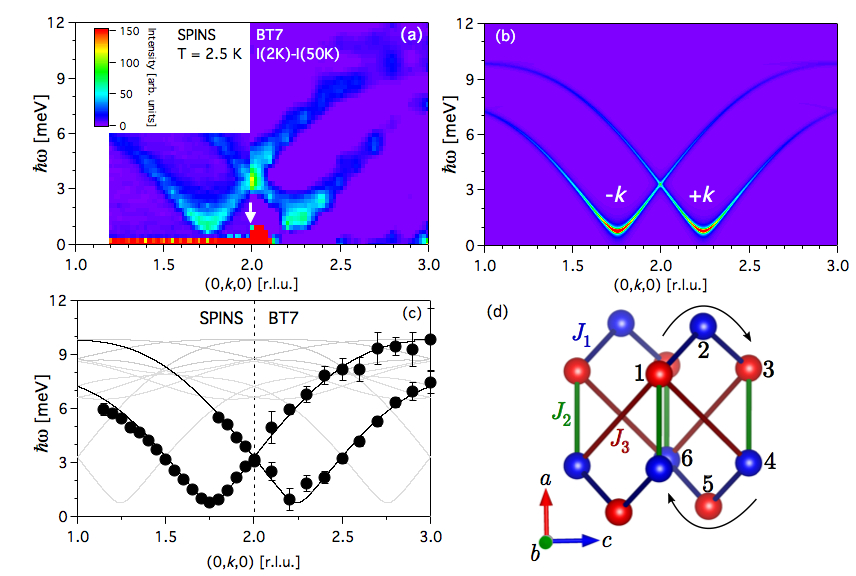}
\caption{The magnetic excitations in $\alpha$-Cu$_{2}$V$_{2}$O$_{7}$. (a)~The contour map, which is constructed from a series of constant-$\mathbf{Q}$ scans taken at SPINS for $k<2$ and at BT7 for $k>2$, shows two dispersive branches of the excitations centred around $(0,1.75,0)$ and $(0,2.25,0)$. The BT7 contour map represent the intensity difference between the data measured at 2~K and 50~K. An arrow denotes the magnetic Bragg reflection at $(0,2,0)$. (b)~The contour map of the calculated dynamical structure factor $S(\mathbf{Q},\hbar\omega)$ representing the spin-wave intensity along $K$ was calculated using the obtained fit parameters. (c)~The spin-wave dispersions of all 16 branches are shown in grey for low intensity branches and black for the most intense branches. The data points were obtained from resolution-convolution fitting of the constant-$\mathbf{Q}$ scans. Error bars denote standard derivations throughout the article. (d)~The spin network of $\alpha$-Cu$_{2}$V$_{2}$O$_{7}$ is formed by three dominant exchange interactions $J_1$, $J_2$ and $J_3$. Red and blue spheres represent two spin sublattices, which are parallel and antiparallel to the $a$-axis, respectively. Arrows and spin labels denote the order of the cross product.}\label{fig1}
\end{center}
\end{figure*}

In the magnetically ordered state, $S=1/2$ Cu$^{2+}$ spins in $\alpha$-Cu$_{2}$V$_{2}$O$_{7}$ align antiparallel along the crystallographic $a$-axis forming a collinear structure. In the presence of a magnetic field along the $c$-axis, weak ferromagnetism resulting from the DM-interaction-induced canted moments was observed. When the magnetic field is applied along the $a$-axis, two magnetic transitions appear: one transition at $\mu_0 H_{c1}={}$6.5~T is characterized as the spin-flop transition whereas the other at $\mu_0 H_{c2} {}={}$18.0~T is a result of the spin-flip~\cite{ganatee2}. Combined density functional theory (DFT) calculations and Quantum Monte Carlo simulations suggest a complex spin-network shown in Fig.~\ref{fig1}(d)~\cite{Sannigrahi,Banerjee,ganatee2}. To the first approximation, the spin Hamiltonian for $\alpha$-Cu$_2$V$_2$O$_7$ can be described by~\cite{motome}
\begin{eqnarray}
{\mathcal{H}}&=&
\sum_{i,j}J_{ij}\vec{S}_i\cdot\vec{S}_j+\sum_{k,l}G_{kl}(S_k^xS_l^x-S_k^yS_l^y-S_k^zS_l^z)\nonumber\\
&&\mbox{}+\sum_{k,l}\vec{D}_{kl}\cdot(\vec{S}_k\times\vec{S}_l)-g_e\mu_B\sum_i\vec{S}_i\cdot\vec{B},\label{spinH}
\end{eqnarray}
where the summation $\sum_{i,j}$ $(\sum_{k,l})$ is taken over the nearest, second-nearest, and third-nearest neighbours (nearest neighbours). The first term represents the isotropic exchange interactions, where $J_1$, $J_2$, and $J_3$ depicted in Fig.~\ref{fig1}(d) are all antiferromagnetic with $J_1\sim J_2<J_3$~\mbox{}\cite{ganatee2}. The second term represents the anisotropic exchange interaction $G_1$, which arises from the multiorbital correlation effect caused by the relativistic spin-orbit coupling and multiorbital hybridization~\cite{motome,Hayami1,Hayami2}. The third term denotes the antisymmetric Dzyaloshinskii-Moriya interactions $\vec{D}_1$, which results from the absence of the inversion centre between the nearest-neighbour spins~\cite{Dzyaloshinsky,Moriya}. The last term represents spins in an external magnetic field, where $g_e=-2$ is the electron spin $g$-factor, $\mu_B$ is the Bohr magneton, and $\mathbf{B}$ is the magnetic field applied parallel to the crystallographic $a$-axis. The anisotropic exchange interaction stabilizes the collinear antiferromagnetic spin structure and introduces an energy gap to the magnon excitations, whereas the $a$-component of the DM vector favours a helical spin structure in the $bc$-plane and determines the incommensurate wavevector of the helical modulation. Due to the competition between these two terms, spin fluctuations of the dynamic state may not be around the static spin structure, giving rise to distinct wavevectors for the ordered structure and the minimum of the dynamical spin-waves.

\begin{figure}
\begin{center}
\includegraphics[width=\columnwidth]{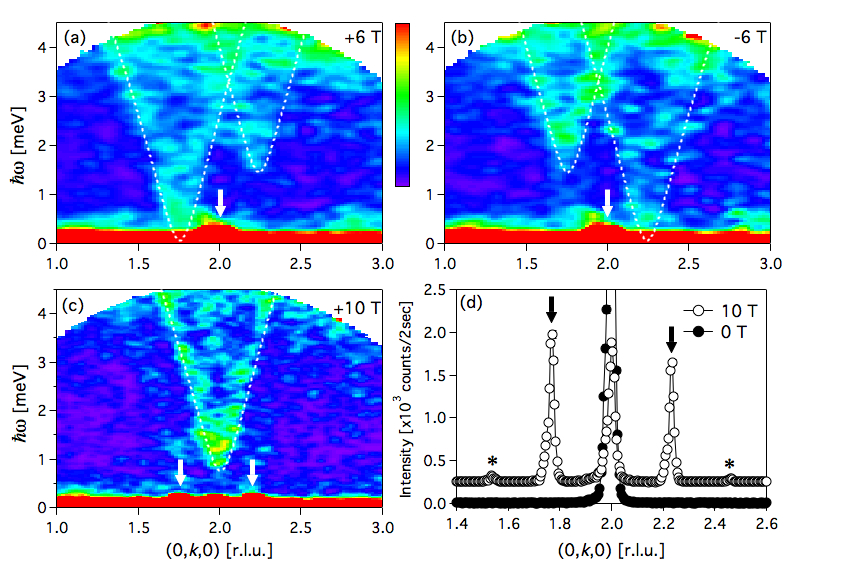}
\caption{Spin-wave excitations in a magnetic field at 1.5 K. (a--c) show the energy-momentum contour maps measured at 6, $-$6, and 10~T, respectively. Dashed lines denote the most intense branches resulting from the spin-wave calculations. (d)~shows the magnetic Bragg peak measured along $(0,k,0)$ below and above the spin-flop transition (similar data measured at different fields are reported in \onlinecite{ganatee2}). Arrows denote the magnetic Bragg peaks, and asterisks ($\ast$) indicate the second harmonic reflections.}\label{fig2}
\end{center}
\end{figure}

Spin dynamics of $\alpha$-Cu$_2$V$_2$O$_7$ was investigated using single-crystal inelastic neutron scattering as described in Methods. An intensity contour map as a function of energy transfer $\hbar\omega$ and momentum transfer $\mathbf{Q}$ measured at the base temperature along $(0,k,0)$ shows two branches of spin-wave excitations symmetrically centred on both sides of the magnetic zone centre $(0,2,0)$, where a magnetic Bragg reflection is observed [Fig.~\ref{fig1}(a)]. The two magnon modes correspond to the clockwise and anticlockwise rotation of spins bonded by $J_1$ with $G_1$ and $\mathbf{D}_1$ along the $[0\bar{1}\bar{1}]$ and $[0\bar{1}1]$ uniform zig-zag chains (Fig.~\ref{figS9}). The degeneracy of these two modes is lifted by the DM interaction, which causes the observed symmetric shift of the magnon modes to the $-k$ (left) and $+k$ (right) side of the zone centre [Fig.~\ref{fig1}(b)]. The $+k$ ($-k$) mode corresponds to the anticlockwise rotation along the $b$-axis but clockwise (anticlockwise) along the spin-chain directions (see Supplementary Material). Constant-energy contour maps of scattering intensity (Fig.~\ref{figS1}) covering a wider range in the $(0kl)$ and $(hk0)$ scattering planes confirm that the nonreciprocity of the magnon dispersion, which was not observed in the recent inelastic neutrons scattering on a powder sample~\cite{Banerjee}, is only along $(0,k,0)$, which is consistent with linear spin-wave calculations [Figs.~\ref{figS1}(g--j)], as will be discussed later. The energy of both modes increases steadily with roughly the same slope up to about 10~meV, and they cross at the zone centre at $\hbar\omega\sim3$~meV. The absence of crossing avoidance suggests that the counter-rotating excitations are decoupled. The energy scan measured at the high-resolution cold neutron spectrometer SPINS yields the gap energy $\Delta = 0.75(6)$~meV at $(0,1.75,0)$ [Fig.~\ref{figS2}(d)].  The gap energy as a function of temperature correlates with the decrease of the order parameter and the peak width becomes broader, which is indicative of shorter life-time as temperature increases toward $T_N$ (Fig.~\ref{figS3}).  These results confirm that the excitations are due to the fluctuations of the ordered magnetic moments.  The nonreciprocal magnons were recently observed in the field-induced ferromagnetic phase of noncentrosymmetric MnSi, where the shift of the  single, non-degenerate magnon mode is asymmetric depending on the field direction~\cite{sato}, in contrast to the symmetric shift in the antiferromagnet at zero field observed in this study.

To quantitatively describe the observed magnon dispersion in $\alpha$-Cu$_2$V$_2$O$_7$, we employed linear spin-wave calculations, which are described in Methods. The calculated magnons consist of 16 modes, denoted by the grey lines in Figs.~\ref{fig1}(c), \ref{figS4}(a), and \ref{figS4}(b). However, only two modes, which were experimentally observed in Fig.~\ref{fig1}(a) and theoretically confirmed in Figs.~\ref{fig1}(b), \ref{figS4}(c), and \ref{figS4}(d), were selected to fit the measured dispersion. The data points in Fig.~\ref{fig1}(c) shows the measured magnon dispersion along $(0,k,0)$ obtained from the constant-$\mathbf{Q}$ scans (Fig.~\ref{figS2}). The dispersion along $(h,1.75,0)$ and $(0,2,l)$ (Fig.~\ref{figS4}) as well as the field-dependence of the energy gap [Fig.~\ref{fig4}(d)] were also measured and used in the global fit to obtain the relevant Hamiltonian parameters in Eq.~\ref{spinH}. The ratio $J_1:J_2:J_3$ was fixed to the result obtained from the DFT calculations of $1.00:1.12:2.03$~\mbox{}\cite{ganatee2}. The minimal model, which includes three isotropic exchange interactions, the anisotropic exchange interaction $G_1$, and the uniform DM vector $\mathbf{D}_1=(D_{1a},0,0)$, is able to capture the magnon nonreciprocity along $(0,k,0)$ and the dispersion as shown by the solid lines in Figs.~\ref{fig1}(c), \ref{figS4}(a), and \ref{figS4}(b). The obtained fitted parameters are $J_1=2.67(1)$~meV, $J_2=2.99$~meV, $J_3=5.42$~meV, $G_1=0.282(1)$~meV, and $D_{1a}=2.79(1)$~meV. The value of the DM parameter with $D_{1a}/J_{1}\sim1$ is much higher than that measured in other $S=1/2$ Cu$^{2+}$ spin systems~\cite{Thio}, suggesting the exceptionally strong spin-orbit coupling.

\begin{figure}
\begin{center}
\includegraphics[width=\columnwidth]{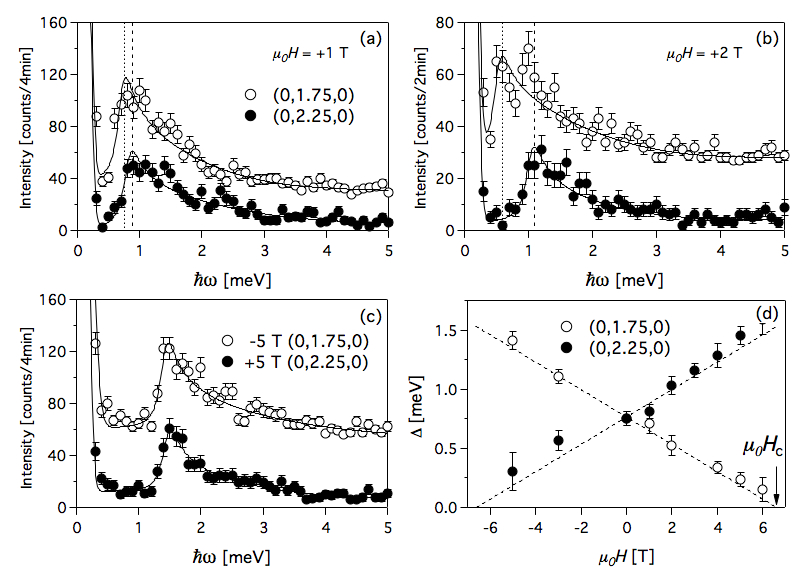}
\caption{Constant-$\mathbf{Q}$ scans were measured at $(0,1.75,0)$ and $(0,2.25,0)$ at (a)~1~T and (b)~2~T. The dashed and dotted lines denote positions of the energy gap. In (c), the scan at (0,1.75,0) and $-5$~T is compared with that at (0,2.25,0) and $+5$~T, which shows the same shift in energy. All (0,1.75,0) data are shifted for clarity. The gap energy as a function of field at $(0,1.75,0)$ and $(0,2.25,0)$ is shown in (d). The critical field is labelled by $\mu_0 H_c$, at which the energy gap vanishes.}
\label{fig3}
\end{center}
\end{figure}

The symmetry between the anticlockwise and clockwise modes is broken in the presence of the applied magnetic field as the electron spins precess under the perpendicular field, and if the spin rotation is in the same (opposite) sense as (to) the spin precession, or anticlockwise (clockwise) rotation, the excitation energy becomes lower (higher) as illustrated for the spin-chain model in Supplementary Material. Experimentally, for $\alpha$-Cu$_2$V$_2$O$_7$ when the applied field is positive along the $a$-axis ($\mu_0 H=+6$~T), the $+k$ mode (clockwise rotation along the zig-zag chain) is shifted upward whereas the $-k$ mode (anticlockwise rotation) is shifted downward [Fig.~\ref{fig2}(a)]; on the other hand, if the field is negative ($\mu_0 H=-6$~T), the shift of the dispersion reverses [Fig.~\ref{fig2}(b)]. Spin-wave calculations confirm the energy shift in the presence of the field as shown in Fig.~\ref{figS5}. The energy scans at $+1$~T ($+2$~T) shown in Figs.~\ref{fig3}(a) [\ref{fig3}(b)] display the asymmetric shift of the gap energy at $(0,1.75,0)$ and $(0,2.25,0)$. Figure~\ref{fig3}(c) depicts the same energy shift of the $+k$ and $-k$ modes when the field of the same magnitude is oppositely aligned. The gap energy as a function of field [Fig.~\ref{fig3}(d)] shows a linear relation, consistent with the spin-chain model (Supplementary Material), with a negative slope for $(0,1.75,0)$ and positive slope for $(0,2.25,0)$. Extrapolating the linear relation to intersect the horizontal axis yields the critical field $\mu_0 H_{c1}$ of $\pm$6.61(2)~T, at which the energy gaps at $(0,2\mp0.25,0)$ close and the spin-flop transition occurs~\cite{ganatee2}.

When the applied field is increased from $+6$~T to $+10$~T, the magnetic Bragg reflection at $(0,2,0)$, indicated by the arrow in Fig.~\ref{fig2}(a), moves to the incommensurate wavevectors $(0,2\pm\delta,0)$ where $\delta\sim0.23$, as denoted by the pair of arrows in Figs.~\ref{fig2}(c) and~\ref{fig2}(d). The transfer of the Bragg intensity documented in Fig.~\ref{fig2}(d) occurs at the spin-flop transition reported at $\mu_0 H_{c1}=6.5$~T~\cite{ganatee2}, and is consistent with the transition from the collinear spin structure to the helical spin structure with the majority of the spin component being in the $bc$-plane. The harmonic reflections at roughly $(0,2\pm2\delta,0)$ indicated by the asterisks in Fig.~\ref{fig2}(d) substantiate the incommensurate modulation of the helical structure. The spin-wave excitations in the spin-flop state show the minimum at $(0,2,0)$ [Fig.~\ref{fig2}(c)] indicative of the reciprocal magnons. Hence, while the collinear spin structure below $\mu_0 H_{c1}$ hosts the nonreciprocal magnons, the helical spin structure above $\mu_0 H_{c1}$ gives rise to the reciprocal magnons with the polarisation most likely along the $a$-axis, highlighting the competitive nature of the anisotropic exchange and antisymmetric DM interactions.

\begin{figure}
\begin{center}
\includegraphics[width=\columnwidth]{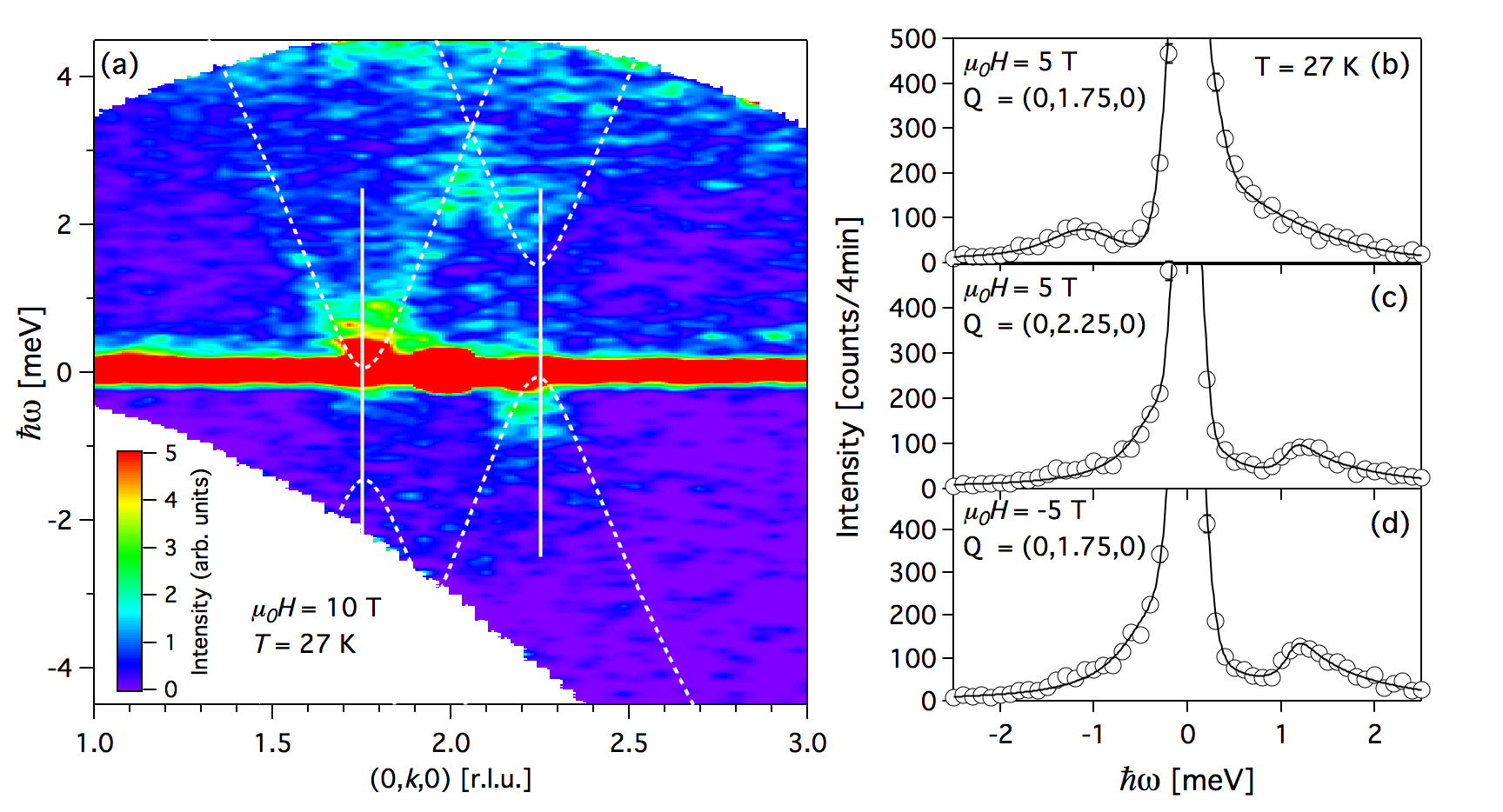}
\caption{Detailed balance relation of the scattering intensity at 27~K. (a)~The energy-momentum contour map of the scattering intensity was measured at 10~T, which puts the system in the co-linear antiferromagnetic state in the phase diagram. The solid lines show the constant-$\mathbf{Q}$ scans, which were performed to investigate the negative-energy-transfer and positive-energy-transfer excitations for (b)~$\mathbf{Q}=(0,1.75,0)$ and $\mu_0 H = +5$~T, (c)~$\mathbf{Q}=(0,2.25,0)$ and $\mu_0 H = +5$~T, and (d) $\mathbf{Q}=(0,1.75,0)$ and $\mu_0 H = -5$~T.}\label{fig4}
\end{center}
\end{figure}

The asymmetry between the $-k$ and $+k$ modes provides a great opportunity to verify the well-known detailed balance relation for the dynamical structure factor, which can be described by~\cite{squires}
\begin{equation}
S(-\mathbf{q},-\hbar\omega)=\e^{-\hbar\omega/k_BT}S(\mathbf{q},\hbar\omega),\;\label{bdrel}
\end{equation}
where $k_B$ is the Boltzmann constant, and the momentum transfer $\mathbf{q}$ is measured from the zone centre. The breaking of inversion symmetry in the $\alpha$-Cu$_2$V$_2$O$_7$ crystal structure and of time-reversal symmetry due to the applied field introduces the asymmetry between the clockwise and anticlockwise magnon modes, with the result that $S(-\mathbf{q},\pm\hbar\omega)\neq S(+\mathbf{q},\pm\hbar\omega)$. The energy-momentum contour map was measured at 27~K and 10~T, where the system is in the collinear antiferromagnetic states as the critical field $\mu_0 H_{c1}$ increases to $\sim15$~T at $T\sim25$~K (Fig.~\ref{figS10}). The elevated temperature is necessary to populate magnons and facilitate the neutron-energy-gain (negative energy transfer) scattering process. Figure~\ref{fig4}(a) illustrates the asymmetry upon the sign reversal of $\mathbf{q}$ and $\hbar\omega$ separately, reflecting broken inversion and time reversal symmetry, respectively. The detailed measurements of the constant-$\mathbf{Q}$ scans extended to negative energy shown in Figs.~\ref{fig4}(b--d) display the asymmetry between the neutron-energy-gain and neutron-energy-loss scattering intensity satisfying the detailed balance relation in Eq.~\ref{bdrel}; the asymmetry is reversed when the momentum transfer changes from $+\mathbf{q}$ to $-\mathbf{q}$ and vice versa [Figs.~\ref{fig4}(b) and \ref{fig4}(c)], and when the field direction is flipped [Figs.~\ref{fig4}(b) and \ref{fig4}(d)]. 

The magnon nonreciprocity due to the DM interaction causes the clockwise and anticlockwise modes to acquire different phase velocities resulting in the rotation of the magnon polarisation, called the spontaneous magnonic Faraday effect.  In contrast to a ferromagnet~\cite{Nembach,Kwon}, the magnonic Faraday rotation in the antiferromagnet is opposite for the left- and right-propagating spin-waves, thus giving rise to the net nonreciprocal phase flow in thermal equilibrium. Taking the role of the DM interaction, an external electric field, which breaks inversion symmetry and lifts the degeneracy of the two counter-rotating modes, can cause magnon nonreciprocity and lead to the electric-field-induced magnonic Faraday effect, which may find applications in the spin-wave field-effect transistor~\cite{Cheng:2016kv}. \\[3mm]

\textbf{Methods}\\[3mm]

Single crystals of $\alpha$-Cu$_2$V$_2$O$_7$ were grown using the method described in Ref.~\onlinecite{ganatee}. Inelastic neutron scattering measurements were conducted to study magnetic excitations in $\alpha$-Cu$_2$V$_2$O$_7$ using the thermal-neutron triple-axis spectrometer BT7~\cite{Lynn2012}, the cold-neutron triple-axis spectrometer SPINS, the Multi Axis Crystal Spectrometer (MACS)~\cite{Rodriguez:2008be}, and the Disk Chopper time-of-flight Spectrometer (DCS)~\cite{Copley2003477}, all of which are located at the NIST Center for Neutron Research (NCNR), Gaithersburg, MD, USA\@. For the BT7 measurements, a single crystal of mass 1.39~g was aligned so that $(0,k,l)$ was in the scattering plane. The final energy was fixed at 14.7~meV with horizontal collimations of open -- 80$'$ -- 80$'$ -- 120$'$. A PG filter was placed after the sample to eliminate the higher order contamination. The crystal was cooled to the base temperature of 2~K using a $^4$He closed-cycle cryostat. A series of constant-$\mathbf{Q}$ scans was collected at the base temperature and at 50~K for background subtraction. The measurements at SPINS were performed using neutrons with a fixed final energy of 5~meV and horizontal collimations of open -- 80$'$ -- 80$'$ -- open, yielding the energy resolution of $\sim0.5$~meV. A cold Be filter was placed in the scattered beam to cut off the neutron energy higher than 5~meV. A series of constant-$\mathbf{Q}$ scans was measured in the $(0kl)$ and $(hk0)$ planes. All scans were measured at the base temperature of 2.5~K. At MACS, contour maps at constant energies were measured in the $(0kl)$ and $(hk0)$ planes. The fixed final energy of 5~meV was selected using the vertically focusing, horizontally flat analyser with cold Be filters in the scattered beam. The base temperature of 1.7~K was achieved using a $^4$He cryostat. In the $(0kl)$ plane, the constant energy cuts were measured between 0 and 12~meV and the background was measured at 40~K and 80~K for $\hbar\omega{}={}$2.5 and 5.0~meV, whereas in the $(hk0)$ plane, the constant energy cuts were measured between 0 and 6~meV and the background was measured at 80~K for $\hbar\omega{}={}$2.0 and 5.0~meV. For the DCS experiment, the single crystal was measured in magnetic fields up to 10~T using neutrons with incident energy $E_i = 5.98$ meV, yielding the instrumental energy resolution of $0.3$~meV at the elastic position. The crystal was mounted inside a vertical-field 10-T magnet and aligned so that the crystallographic $a$-axis was perpendicular to the scattering plane and parallel to the applied field. The magnet was equipped with a closed-cycle $^4$He cryostat, which can reach the  base temperature of 1.5~K.

Linear spin-wave calculations were performed based on the collinear, antiferromagnetic spin structure and on the spin Hamiltonian in Eq.~\ref{spinH} using the standard Holstein-Primakoff transformation and Fourier transform implemented using a symbolic algebraic method. We choose $\hat{x}$, $\hat{y}$, and $\hat{z}$ to be along the crystallographic $a$-, $b$-, and $c$-axis, respectively. There are a total of 16 spins in the unit cell and hence 16 spin-wave modes as shown by the grey lines in Figs.~\ref{fig1}(c), \ref{figS4}(a), and \ref{figS4}(b). The spin-wave energies are  non-negative eigenvalues obtained from diagonalising a $32\times32$ matrix. The spin-wave intensity for each mode was calculated using~\cite{squires}
\begin{equation*}
I(\mathbf{Q},\hbar\omega)=\left(\frac{\gamma
r_0}{2}\right)^2f(\mathbf{Q})^2\sum_{\alpha,\beta}
(\delta_{\alpha\beta}-\hat{Q}_\alpha\hat{Q}_\beta)S^{\alpha\beta}(\mathbf{Q},\hbar\omega),\;\label{eqInt}
\end{equation*}
where $(\gamma r_0/2)^2=72.65\times10^{-3}$ barn $/\mu_B^2$, and $f(\mathbf{Q})$ is the magnetic form factor for Cu$^{2+}$ ions. $\sum_{\alpha,\beta} (\delta_{\alpha\beta}-\hat{Q}_\alpha\hat{Q}_\beta)$ is the geometric factor, where $\alpha$ and $\beta$ are vector components $\hat{x}$, $\hat{y}$, or $\hat{z}$. $S^{\alpha\beta}(\mathbf{Q},\hbar\omega)$ is the dynamical structure factor, which is the spatial and temporal Fourier transform of the spin-pair correlation function $\braket{S^{\alpha}(0,0)S^{\beta}(\mathbf{r},t)}$ that can be calculated from the eigenvectors.\\[3mm]

\textbf{Acknowledgements}\\[3mm]
We thank Y. Motome for fruitful discussion. Work at Mahidol University was supported in part by the Thailand Research Fund Grant Number RSA5880037 and the Thailand Centre of Excellence in Physics.  Work at NCNR was supported in part by the National Science Foundation under Agreement No. DMR-1508249. Work at IMRAM was partly supported by a Grant-In-Aid for Scientific Research (24224009) from the Japan Society for the Promotion of Science (JSPS), and by the Research Program ``Dynamic Alliance for Open Innovation Bridging Human, Environment and Materials".\\[3mm]

\textbf{Author Contributions}\\[5mm] G.G., Y.Z., Y.Q., L.W.H., N.P.B., and K.M. performed the neutron scattering experiments. P.P., T.J.S., and K.M. performed the linear spin-wave calculations. G.G., P.P., and K.M. wrote the manuscript. All authors commented on the final version of the manuscript.\\[3mm]

\textbf{Competing Interests} The authors declare that they have no competing financial interests.\\[3mm]

\textbf{Correspondence} Correspondence and requests for materials should be addressed to K. Matan~(kittiwit.mat@mahidol.ac.th).\\[3mm]


\clearpage
\pagebreak
\widetext
\begin{center}
\textbf{\large Supplementary Material\\Nonreciprocal magnons and
symmetry-breaking in the noncentrosymmetric antiferromagnet}
\end{center}
\setcounter{equation}{0}
\setcounter{figure}{0}
\setcounter{table}{0}
\setcounter{page}{1}
\makeatletter
\renewcommand{\theequation}{S\arabic{equation}}
\renewcommand{\thefigure}{S\arabic{figure}}

\section{Contour maps in $(0kl)$ and $(hk0)$ planes}
\begin{figure}[htp]
\begin{center}
\includegraphics[height=10cm]{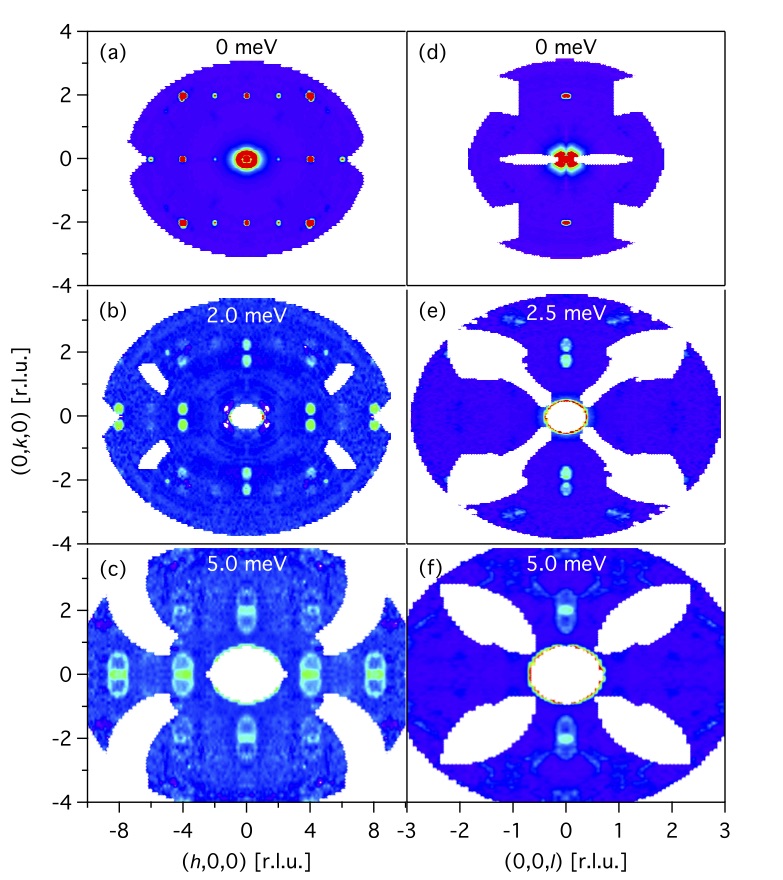}
\includegraphics[height=7cm]{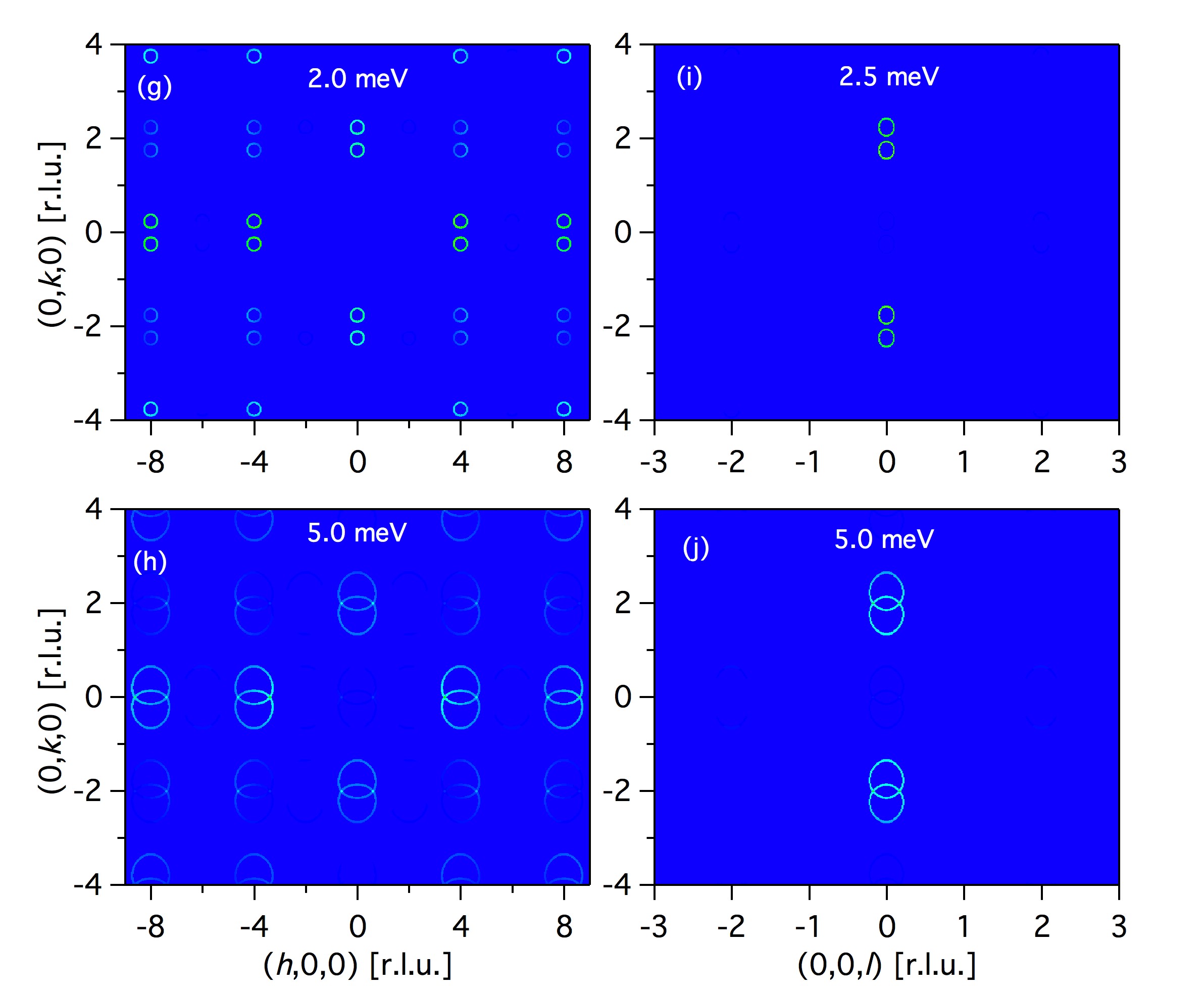}
\caption{Constant-energy contour maps were measured using MACS at 1.5~K in the $(hk0)$-plane at (a)~zero energy, (b)~2.0~meV, and (c)~5.0~meV. Another set of contour maps were measured in the $(0kl)$-plane at (d)~zero energy, (e)~2.5~meV, and (f)~5.0~meV. The data shown in (b) and (c)~are background subtracted by the background that was measured at 80~K. The data in the $(hk0)$-plane ($(0kl)$-plane) are folded with respect to $[1,0,0]$ and $[0,1,0]$ ($[0,1,0]$ and $[0,0,1]$). Contour maps of the calculated scattering intensity are shown for the $(hk0)$ plane at (g)~2.0~meV and (h)~5.0~meV, and for the $(0kl)$ plane at (i)~2.5~meV and (j)~5.0~meV.}\label{figS1}
\end{center}
\end{figure}

Constant-energy contour maps of scattering intensity as a function of momentum transfer were measured using MACS covering a wide $\mathbf{Q}$ range in the $(0kl)$ and $(hk0)$ scattering planes to study the splitting of the excitations around the magnetic zone centres. Figs.~\ref{figS1}(a--c) [\ref{figS1}(d--f)] show the folded contour maps at zero energy, 2.0 (2.5)~meV, and 5.0~meV in the $(hk0)$ [$(0kl)$] plane. At $\hbar\omega{}={}$0~meV [Figs.~\ref{figS1}(a) and~\ref{figS1}(d)], the magnetic Bragg reflections, which are on top of the nuclear Bragg peaks, can be observed as a very intense single spot at all expected magnetic zone centres. At finite energy transfers, the contour maps show the splitting of the minimum point of the magnon dispersion around the zone centres along the $k$ direction. This splitting can be observed above the gap energy at $\hbar\omega{}={}$2.0~meV for the $(hk0)$ plane [Fig.~\ref{figS1}(b)] and 2.5~meV for the $(0kl)$ plane [Fig.~\ref{figS1}(e)] at all magnetic zone centres. At higher energies, the excitations become dispersive and form two ellipses which start to overlap at $\hbar\omega\sim3$~meV. The ellipses become larger and more elongated along $k$ with increasing energy transfer as shown in Figs.~\ref{figS1}(c) and \ref{figS1}(f) for $\hbar\omega=5.0$~meV. For comparison, the calculated scattering intensity maps plotted in Figs.~\ref{figS1}(g--j) show good agreement with the measured contour maps at the corresponding energies. The calculated intensity is integrated using an energy width of $0.5$ meV with the mid-point at the labelled energy. The discrepancy between the measured and calculated results, i.e., the filled ovals for the measured data versus open circles for the calculated results, is due to the fact that the instrumental-resolution effect was ignored in the calculations.

\section{Constant-$\mathbf{Q}$ scans}
\begin{figure}[htp]
\begin{center}
\includegraphics[height=15cm]{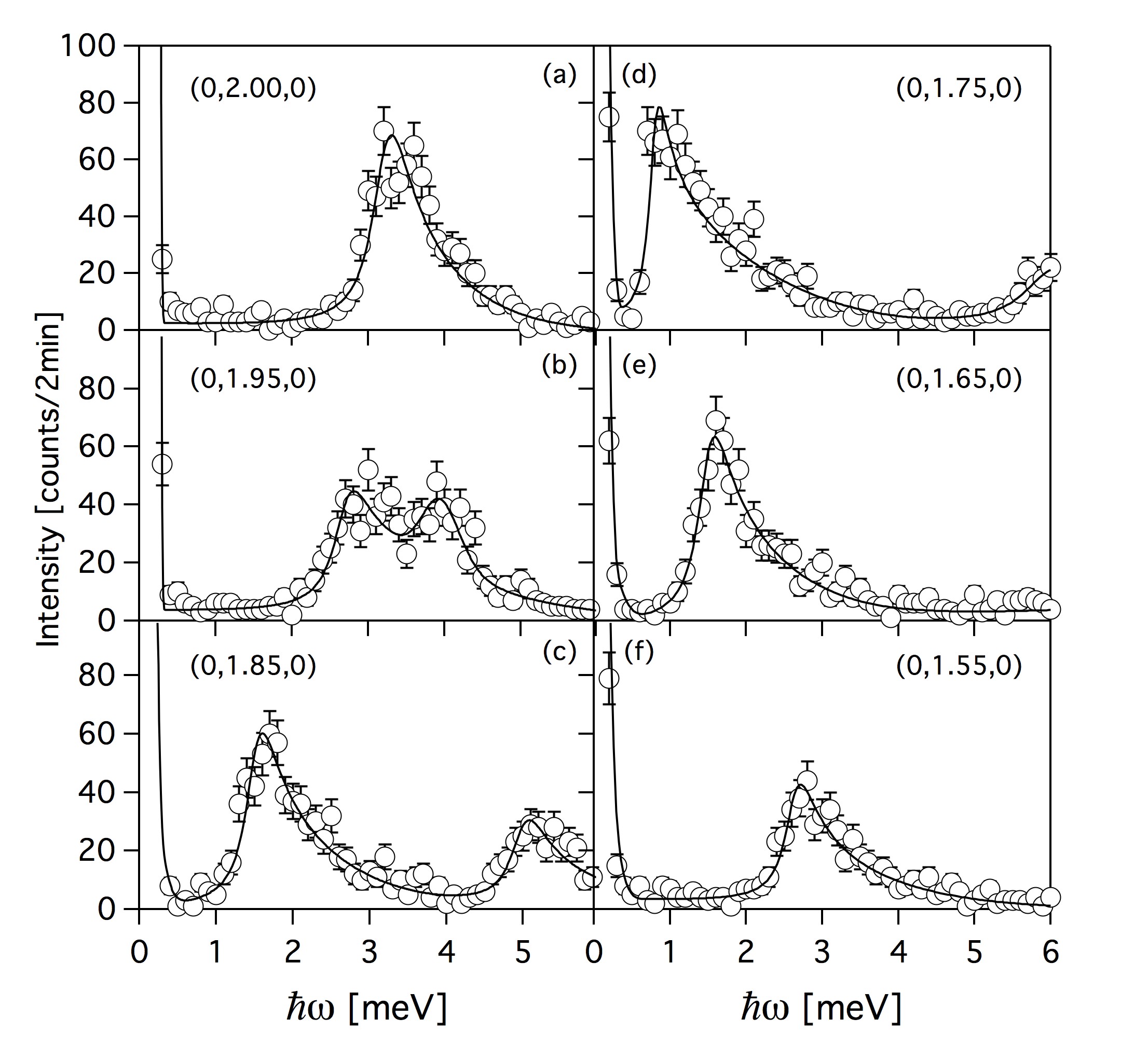}
\caption{Representative constant-$\mathbf{Q}$ scans measured on SPINS were fitted to the empirical dispersion relation convoluted with the resolution function. The line-shape of the observed peak is governed by the convolution with the instrumental resolution, indicative of the resolution-limited peak width.}\label{figS2}
\end{center}
\end{figure}

All constant-$\mathbf{Q}$ scans, which were used to construct the contour maps in Fig.~\ref{fig1}(a), were fitted using an empirical dispersion convoluted with the instrumental resolution function in order to extract the excitation energies as a function of $\mathbf{Q}$. Representative energy scans at several $\mathbf{Q}$'s around $(0,2,0)$ measured at SPINS are shown with the convoluted fit in Fig.~\ref{figS2}. The fitting results including those obtained from fitting the BT7 data are plotted in Fig.~\ref{fig1}(c).

\newpage
\section{Temperature dependence of gap energy}

\begin{figure}[htp]
\begin{center}
\includegraphics[height=15cm]{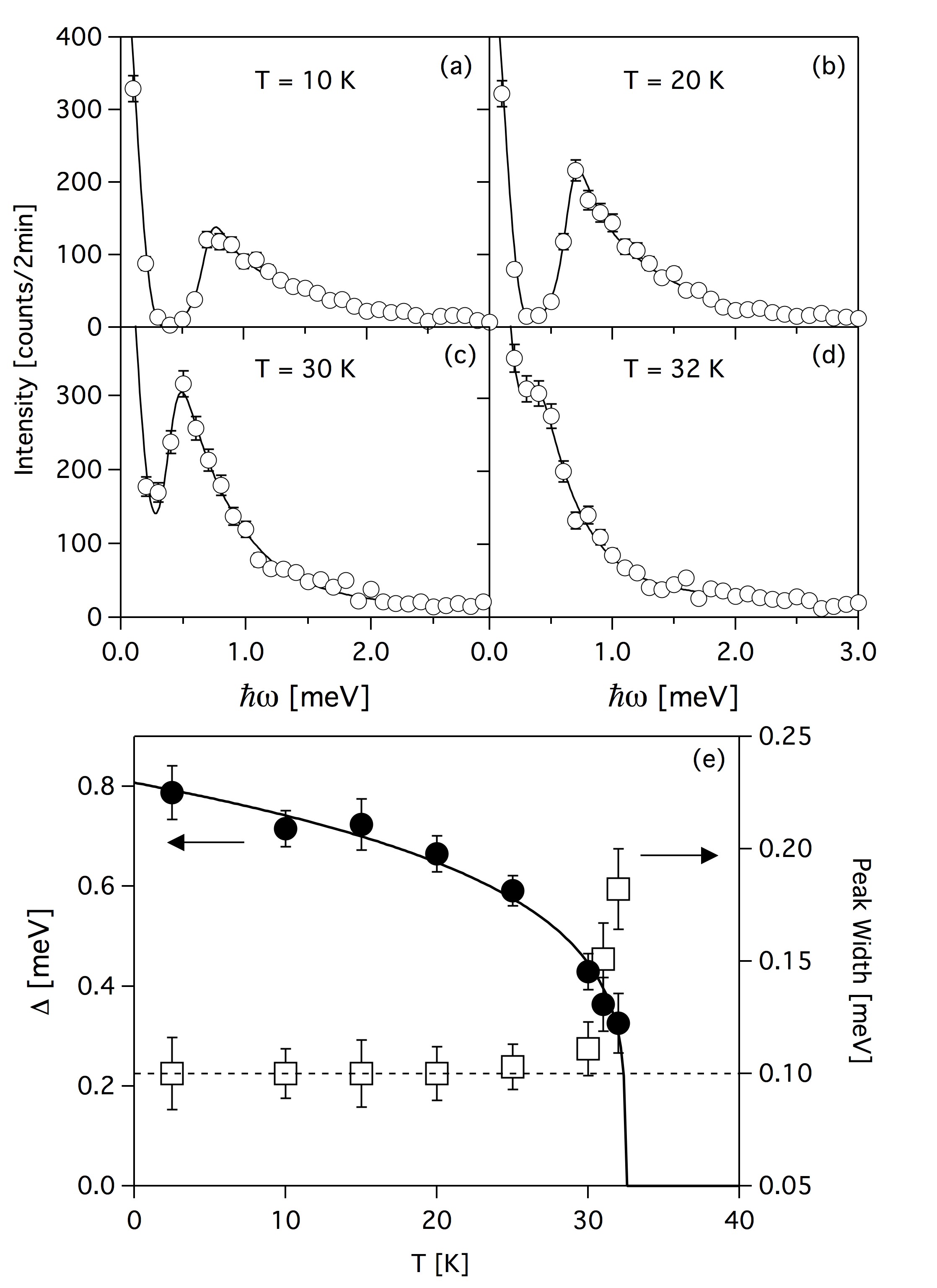}
\caption{Temperature dependence of the gap energy $\Delta$ at $(0,1.75,0)$. (a--d) show constant-$\mathbf{Q}$ scans at four temperatures that are fitted to the empirical dispersion convoluted with the resolution function. (e)~displays the gap energy $\Delta$ (closed symbols) and peak width (open symbols) as a function of temperature. The solid line denotes a fit the power law described in the text. The dashed line represents the instrumental resolution.}
\label{figS3}
\end{center}
\end{figure}

\begin{figure}[htp]
\begin{center}
\includegraphics[height=12cm]{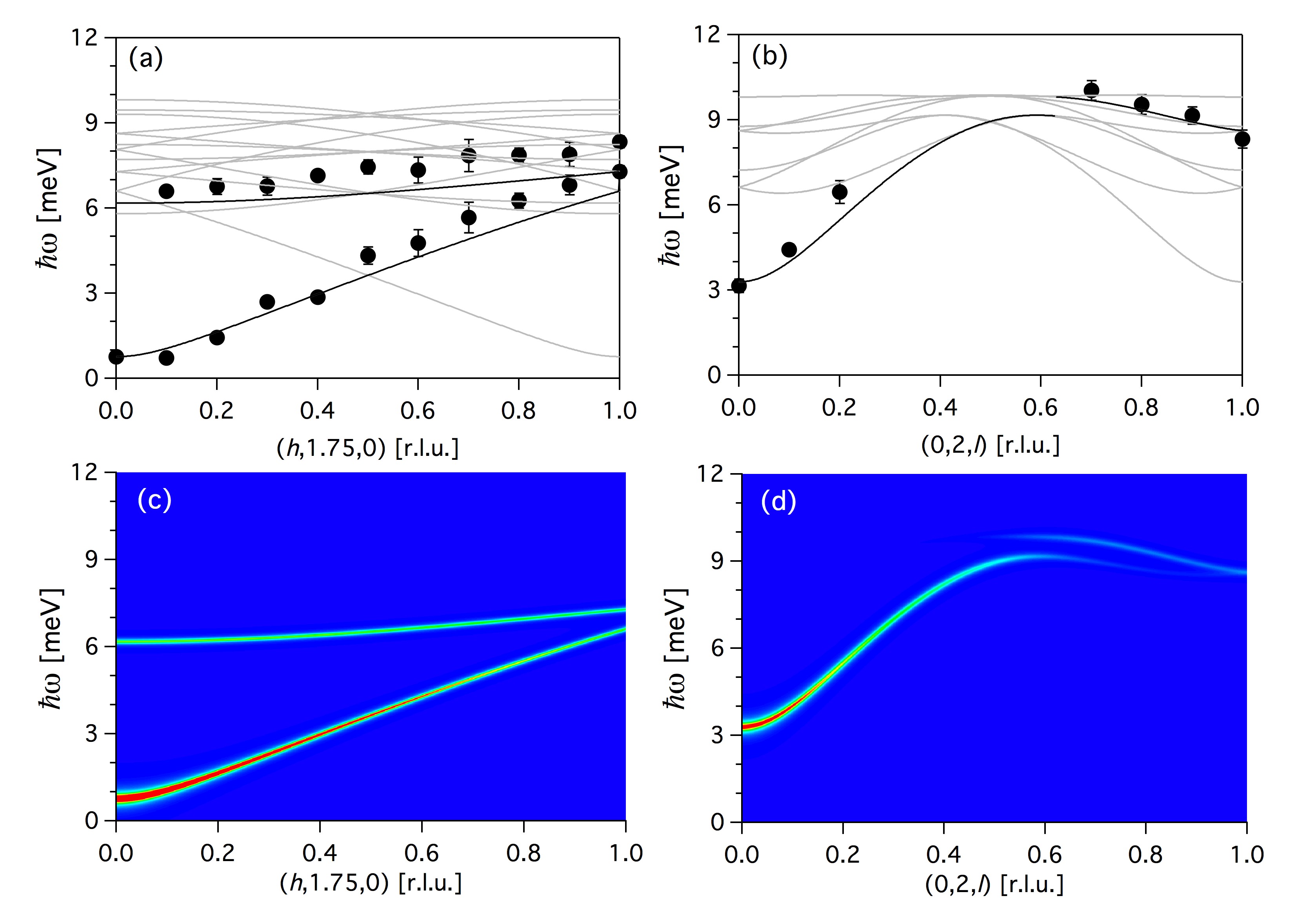}
\caption{Dispersion along (a)~$(h, 1.75,0)$ and (b)~$(0,2,l)$. Solid lines show the high intensity modes, which were used to fit the data. (c) and (d)~show the calculated intensity contour maps depicting the high-intensity modes.}\label{figS4}
\end{center}
\end{figure}

\begin{figure}[htp]
\begin{center}
\includegraphics[height=6cm]{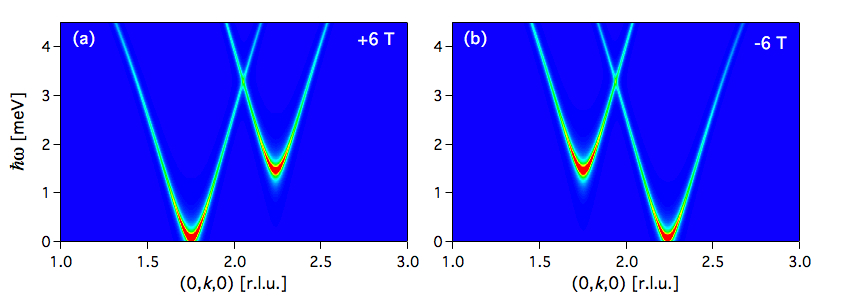}
\caption{Calculated intensity contour maps show the magnon dispersion in the presence of the applied magnetic field of (a)~6~T and (b)~$-6$~T. All spin Hamiltonian parameters are fixed to the values obtained from the global fit.}\label{figS5}
\end{center}
\end{figure}

The constant-$\mathbf{Q}$ scans at $(0,1.75,0)$ were performed at several temperatures below $T_N$ to study the evolution of the gap energy $\Delta$ as the temperature approaches $T_N$. The data were fitted and convoluted with the instrumental resolution function as shown by solid lines in Figs.~\ref{figS5}(a--d). The obtained fitted gap energies were then plotted as a function of temperature in Fig.~\ref{figS3}(e). The results exhibit a decrease of the gap energy, which correlates with the decline of the order parameter as $T_N$ is approached. The fit to the power law $\Delta\propto(T_N-T)^{\beta}$ yields $T_N=32.5(6)$~K, in agreement with $T_N$ obtained from the order parameter fit, and $\beta=0.23(5)$, consistent with the exponent describing the ordered moment $M$, which scales with the square root of the magnetic Bragg scattering intensity $(M\propto \sqrt{I})$. This result confirms that the magnetic excitations are due to the fluctuations of the ordered moments. Furthermore, as the temperature increases toward $T_N$, the peak width becomes broader, indicative of a shorter life-time of the excitations typical for the spin-waves.

\clearpage
\section{Nonreciprocal magnons in an antiferromagnetic spin-chain}
To illustrate the origin of the nonreciprocal magnons in $\alpha$-Cu$_2$V$_2$O$_7$, we consider a simple model of an antiferromagnetic spin chain, where spin-waves can be analytically calculated. We hope that this calculation will provide the reader with some insight, which is probably lost in the complicated spin network of $\alpha$-Cu$_2$V$_2$O$_7$ and the numerical calculations, into the underlying mechanism giving rise to the nonreciprocal magnons. We note that when only the nearest-neighbour spins connected by $J_1$ are considered, Cu$^{2+}$ spins form a uniform zig-zag chain along the $[0\bar{1}\bar{1}]$ and $[0\bar{1}1]$ directions (Fig.~\ref{figS9}), and hence the spin-chain model should qualitatively and sufficiently describe the magnon dispersion in $\alpha$-Cu$_2$V$_2$O$_7$. The Dzyaloshinskii-Moriya (DM) interactions and anisotropic exchange interactions are introduced to the two-sublattice Heisenberg antiferromagnetic spin-chain model in order to study their effects on the elementary excitations of the system. The calculations show that the gap in the spin-wave spectrum is the result of the anisotropic exchange interactions and the nonreciprocity of
the spin-wave spectrum is due to the DM interactions.

\subsection*{Spin-wave calculations}
\begin{figure}[!hbt]
\begin{center}
\includegraphics[height=3cm]{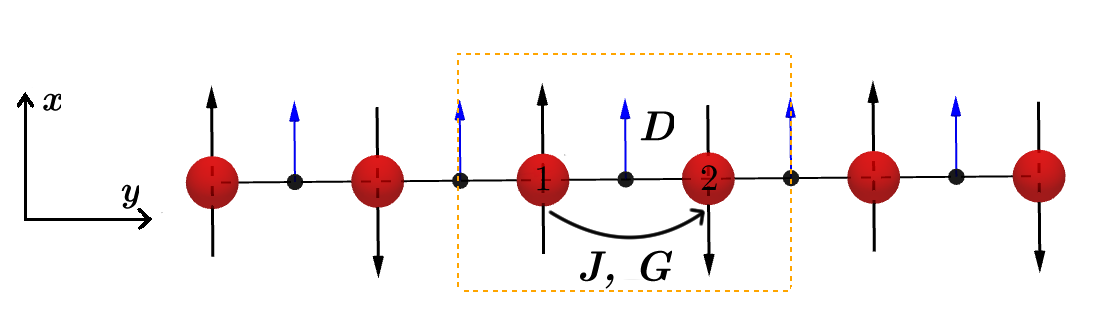}
\caption{The antiferromagnetic spin-chain model comprises the uniform DM
interactions $D$ that point along the $+x$ direction, exchange interaction
$J$, and anisotropic exchange interaction interaction $G$ between the nearest
neighbors. A unit cell with two spin sublattices labelled by 1 and 2 is
enclosed by the orange dashed rectangle.}
\label{figS6}
\end{center}
\end{figure}

As with the Hamiltonian for $\alpha$-Cu$_2$V$_2$O$_7$, the Hamiltonian for this spin-chain model is given by
\begin{equation}
{\mathcal{H}}=\sum_{i,j}J_{ij}\mathbf{S}_i\cdot\mathbf{S}_j+\sum_{i,j}G_{i,j}(S_i^xS_j^x-S_i^yS_j^y-S_i^zS_j^z)
\mbox{}+\mathbf{D}_{i,j}\cdot(\mathbf{S}_i\times\mathbf{S}_j)-g_e\mu_B\sum_i\mathbf{S}_i\cdot\mathbf{B},
\end{equation}
where the summation $\sum_{i,j}$ between spin operators is taken over the nearest-neighbor bonds (Fig.~\ref{figS6}). The first and second terms represent the isotropic exchange interactions $J$ and anisotropic exchange interactions $G$, respectively. The third term is the antisymmetric DM interactions, and the last term represents the interactions between the spins and the external magnetic field $\mathbf{B}=(B,0,0)$.  We note that the one-dimensional lattice spacing is equal to one.

\begin{figure}[!hbt]
\begin{center}
\includegraphics[height=5cm]{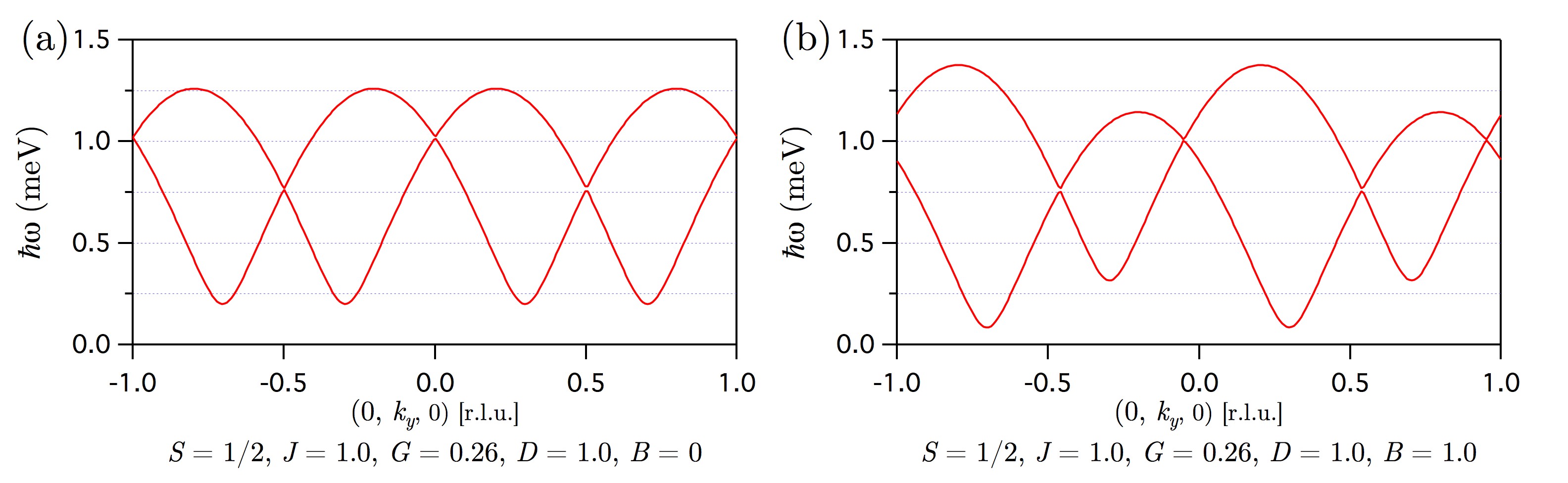}
\caption{The spin-wave spectrum of the antiferromagnet spin chain model for
(a)~$B=0$ and (b)~$B=1$~T.}
\label{figS7}
\end{center}
\end{figure}

We transform coordinates from the spin local coordinates (primed coordinates), in which the positive $z'$-axis coincides with the spin direction, to the crystal coordinates (unprimed coordinates) using the following transformation matrices
\[
\tilde{R}_1=
\begin{pmatrix}
0 & 0 & 1 \\
0 & 1 & 0 \\
-1 & 0 & 0
\end{pmatrix}\textrm{~~~and~~~}
\tilde{R}_2=
\begin{pmatrix}
0 & 0 & -1 \\
0 & 1 & 0 \\
1 & 0 & 0
\end{pmatrix}.
\]
Using the Holstein-Primakoff transformation, the spin operators are transformed to the boson operators $a$ and $a^{\dag}$ via
\begin{equation}
S^{x'}_j=\sqrt{\frac{S}{2}}(a_j+a^{\dag}_j),\quad
S^{y'}_j=\ii\sqrt{\frac{S}{2}}(a_j^{\dag}-a_j),\quad
S^{z'}=S-a_j^{\dag}a_j,\;\label{eqs3}
\end{equation}
where $j=1,2$ for the two sublattices. The spin Hamiltonian can then be written in terms of $a$ and $a^{\dag}$ which now can be solved by introducing the Fourier transform
\begin{equation}
a_\mathbf{k}^j=\sum_\mathbf{k} \e^{-\ii\mathbf{k}\cdot\mathbf{R}_j}a_j, \quad
\mathbf{k}=(0,k_y,0),\;\label{eqs4}
\end{equation}
where the summation runs from $\mathbf{-k}$ to $\mathbf{k}$. Using $(a_\mathbf{k}^1,a_\mathbf{k}^2,a_\mathbf{-k}^{1\dag}, a_\mathbf{-k}^{2\dag})$ as a basis and noting that $a_\mathbf{k}^1\neq a_\mathbf{-k}^1$ for a \textit{noncentrosymmetric} system, we can write the Hamiltonian as
{\small
\[
H=\begin{pmatrix}
(G+J)-g_e\mu_BB & 0 & 0 &
\Delta_{GJ}\cos\frac{k_y}{2}-D\sin\frac{k_y}{2} \\
0 & (G+J)+g_e\mu_BB &
\Delta_{GJ}\cos\frac{k_y}{2}+D\sin\frac{k_y}{2} & 0
\\
0 &
-\Delta_{GJ}\cos\frac{k_y}{2}-D\sin\frac{k_y}{2}
& (-G-J)+g_e\mu_BB & 0 \\
-\Delta_{GJ}\cos\frac{k_y}{2}+D\sin\frac{k_y}{2} &
0
& 0 & (-G-J)-g_e\mu_BB
\end{pmatrix}.
\]
}
We then proceed to solve for the eigenvalues and eigenvectors of this matrix. The eigenvalues of this matrix are
\begin{align*}
E_{val}^1&=A^+-g_e\mu_BB,\quad E_{val}^2=A^-+g_e\mu_BB,\quad
E_{val}^3=-A^-+g_e\mu_BB,\quad E_{val}^4=-A^+-g_e\mu_BB\\
A^{\pm}&=\sqrt{\frac{-D^2+G^2+6GJ+J^2+(D^2-\Delta_{GJ}^2)\cos k_y\pm
2D\Delta_{GJ}\sin k_y}{2}},\\
\Delta_{GJ}&=G-J,
\end{align*}
with the corresponding unnormalized eigenvectors
{\small
\begin{align*}
E_{vec}^1&=\left(\frac{-2G-2J-\sqrt{2}\sqrt{-D^2+G^2+6GJ+J^2+(D^2-\Delta_{GJ}^2)\cos k_y+2D\Delta_{GJ}\sin k_y}}{2\left(\Delta_{GJ}\cos\frac{k_y}{2}-D\sin\frac{k_y}{2}\right)},0,0,1\right)\\
E_{vec}^2&=\left(0,\frac{-2G-2J-\sqrt{2}\sqrt{-D^2+G^2+6GJ+J^2+(D^2-\Delta_{GJ}^2)\cos k_y-2D\Delta_{GJ}\sin k_y}}{2\left(\Delta_{GJ}\cos\frac{k_y}{2}+D\sin\frac{k_y}{2}\right)},1,0\right)\\
E_{vec}^3&=\left(0,\frac{-2G-2J+\sqrt{2}\sqrt{-D^2+G^2+6GJ+J^2+(D^2-\Delta_{GJ}^2)\cos k_y-2D\Delta_{GJ}\sin k_y}}{2\left(\Delta_{GJ}\cos\frac{k_y}{2}+D\sin\frac{k_y}{2}\right)},1,0\right)\\
E_{vec}^4&=\left(\frac{-2G-2J+\sqrt{2}\sqrt{-D^2+G^2+6GJ+J^2+(D^2-\Delta_{GJ}^2)\cos k_y+2D\Delta_{GJ}\sin k_y}}{2\left(\Delta_{GJ}\cos\frac{k_y}{2}-D\sin\frac{k_y}{2}\right)},0,0,1\right).
\end{align*}}
Hence the following pairs of the operators ($a_\mathbf{k}^1$, $a_\mathbf{-k}^{2\dag}$), ($a_\mathbf{k}^2$, $a_\mathbf{-k}^{1\dag}$), ($a_\mathbf{-k}^{1\dag}$, $a_\mathbf{k}^2$), and ($a_\mathbf{-k}^{2\dag}$, $a_\mathbf{k}^1$) are mixed, and we can define the superpositions of these pairs as new boson operators $Y_0(k)$, $Y_1(k)$, $Y_1^{\dag}(-k)$, and $Y_0^{\dag}(-k)$; this procedure is equivalent to the Bogoliubov transformation. The eigenvalues are physically meaningful if they are real, i.e., the term inside the square root of $A^{\pm}$ is positive. The dispersion as a function of momentum transfer along $(0,k,0)$ is shown in Fig.~\ref{figS7}(a).  We note that the eigenvalues are linearly dependent on the magnetic field $B$, the effect of which is shown in Fig.~\ref{figS7}(b).

When the parameters $G$ and $D$ are zero, we obtain
\[
A^{\pm}_J=J\sqrt{\frac{1-\cos k_y}{2}}=J\left|\sin\frac{k_y}{2}\right| ,
\]
which corresponds to the normal antiferromagnetic spin-wave spectrum, where the clockwise and anticlockwise modes are degenerate. Adding $G$ will open the gap (which can be seen from $G^2$ and $6GJ$ in $A^{\pm}$) but does not lift the degeneracy:
\[
A^{\pm}_{JG}=\sqrt{\frac{G^2+6GJ+J^2-\Delta_{GJ}^2\cos k_y}{2}},
\]
and when $D$ is present, the degeneracy is lifted and the spin-wave spectrum will be shifted, in this case, along $k_y$ as shown in Fig.~\ref{figS7}.

\subsection{Spin dynamics}
\begin{figure}[!hbt]
\begin{center}
\includegraphics[height=8cm]{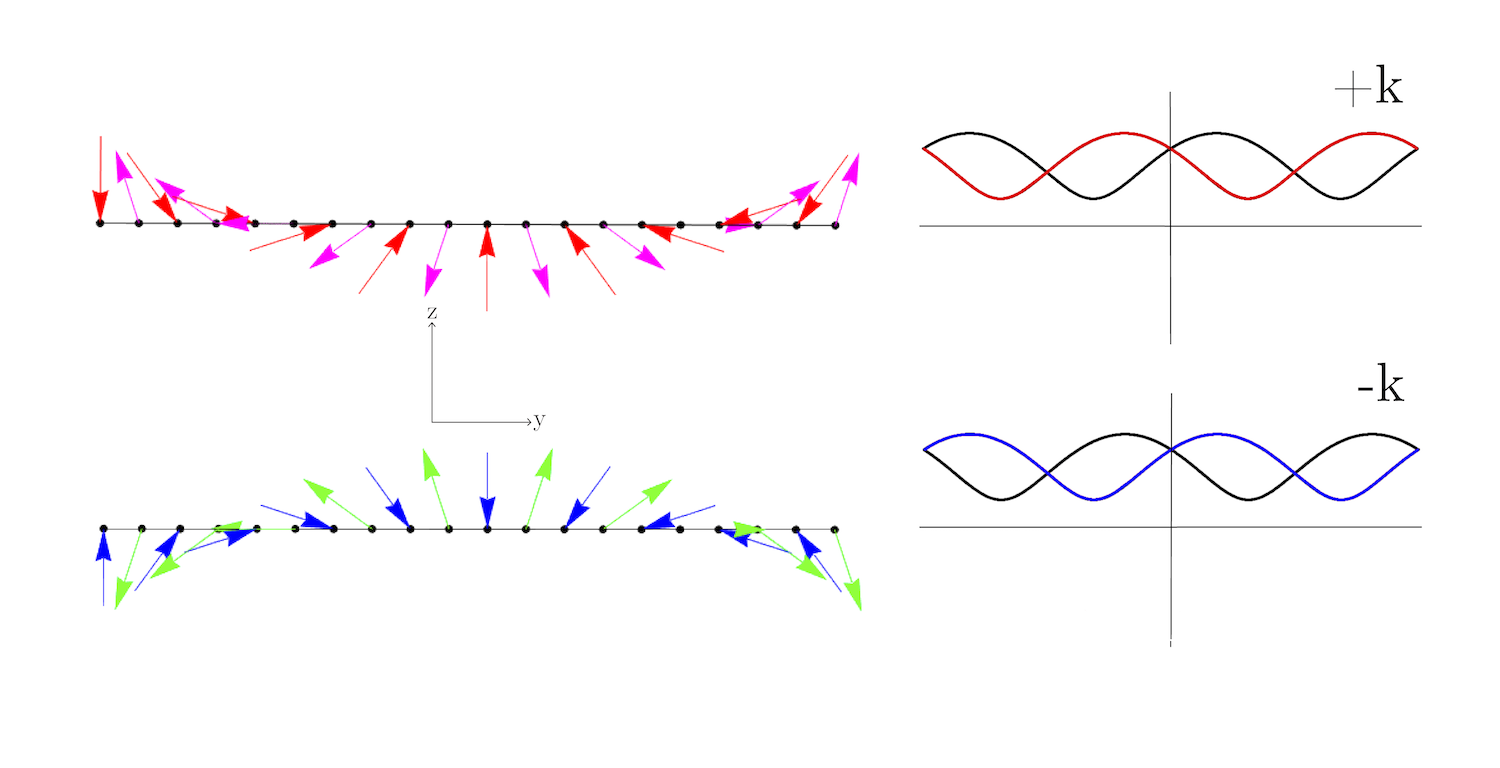}
\caption{Dynamical spin structures are shown in the crystal coordinates on the left column with the corresponding excitation modes on the right column. Top, the structure for the $+k$ mode, where red and magenta arrows denote the sublattice 1 and sublattice 2, respectively. Bottom, the structure for the $-k$ mode, where blue and green arrows denote sublattice 1 and sublattice 2, respectively. Spins for the $+k$ mode rotate anticlockwise whereas the spins for the $-k$ mode rotate clockwise. We note that the chain direction is horizontal from left to right ($+y$ direction), and that only the fluctuating spin component in the $yz$-plane is drawn.}
\label{figS8}
\end{center}
\end{figure}

We can calculate the spin component of the magnon excitations in the local spin coordinates for a particular wavevector $\mathbf{k}$. From Eqs.~\ref{eqs3} and \ref{eqs4}, we obtain
\begin{align*}
S_{x',l}^j&=\sqrt{\frac{S}{2}}(a^j_{\mathbf{k},l}\e^{\ii\mathbf{k}\cdot\mathbf{R}_j}+a^{j\dag}_{\mathbf{k},l}\e^{-\ii\mathbf{k}\cdot\mathbf{R}_j}),\\
S_{y',l}^j&=-\ii\sqrt{\frac{S}{2}}(a^j_{\mathbf{k},l}\e^{\ii\mathbf{k}\cdot\mathbf{R}_j}-a^{j\dag}_{\mathbf{k},l}\e^{-\ii\mathbf{k}\cdot\mathbf{R}_j}),
\end{align*}
where $l$ denotes magnon modes, and $l=1,2$ for the spin-chain model and $l=1,\ldots,16$ for $\alpha$-Cu$_2$V$_2$O$_7$. Using the values of $a$ and $a^{\dag}$ obtained from the eigenvectors, we can calculate and draw dynamical spin structures of the magnon excitations. We found that for the spin-chain model the $+k$ mode is the anticlockwise spin excitation, and the $-k$ mode is the clockwise excitations. Figure~\ref{figS8} shows the anticlockwise and clockwise spin rotations corresponding to the $+k$ mode and the $-k$ mode, respectively, for $k=(0,0.1,0)$ r.l.u., $J=1$, $G=0.005$, and $D=0.1$. When the magnetic field is present, the spins in the $+k (-k)$ mode rotate in the same (opposite) sense as the precession of an electron spin under a perpendicular magnetic field and thus require less (more) energy to excite one magnon [Fig.~\ref{figS7} (b)].

\section{Spin dynamics in $\alpha$-C{\tiny u}$_2$V$_2$O$_7$}
\begin{figure}[!hbt]
\begin{center}
\includegraphics[height=13cm]{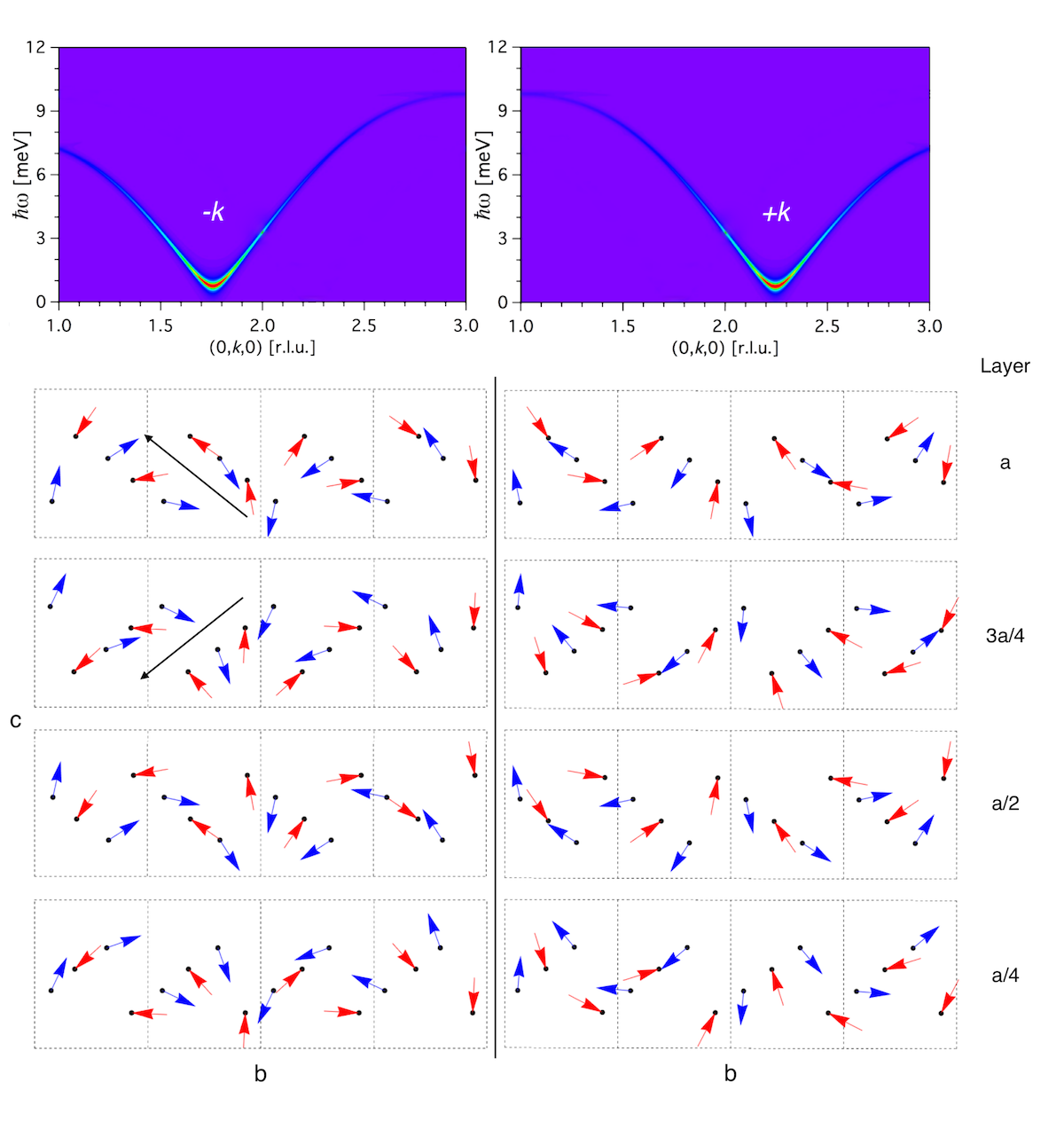}
\caption{Spin dynamics of $\alpha$-Cu$_2$V$_2$O$_7$. The crystallographic $b$-axis is horizontal and positive to the right. The $a$ axis and applied field (positive field) point out of the page. The left (right) panels show the spin dynamics of the $+k$ ($-k$) mode, where the spin rotation is clockwise (anticlockwise) along the spin chain but anticlockwise (clockwise) along the $b$-axis, which is why the shift is to the right (left) of the zone centre. We only draw the $bc$-plane component of the fluctuating spins; the majority of the spin component is anti-aligned and points along $\pm a$. The red and blue arrows denote the spins with the $a$-component along $-a$ and $+a$, respectively.  The spin dynamics on each of the 4 layers in the unit cell, which is enclosed by the dotted line, is drawn separately to provide clearer images of the spin rotation. The spins in 4 unit cells along $b$ are plotted as because of our choice of $\mathbf{q}=(0,\pm0.25,0)$, the dynamic spin structure repeats every 4 unit cells along $b$. Black arrows denote the chain directions, which are along $[0\bar{1}\bar{1}]$ for layers $a/4$ and $3a/4$, and $[0\bar{1}1]$ for layers $a/2$ and $a$.}
\label{figS9}
\end{center}
\end{figure}

\begin{figure}[!hbt]
\begin{center}
\includegraphics[height=10cm]{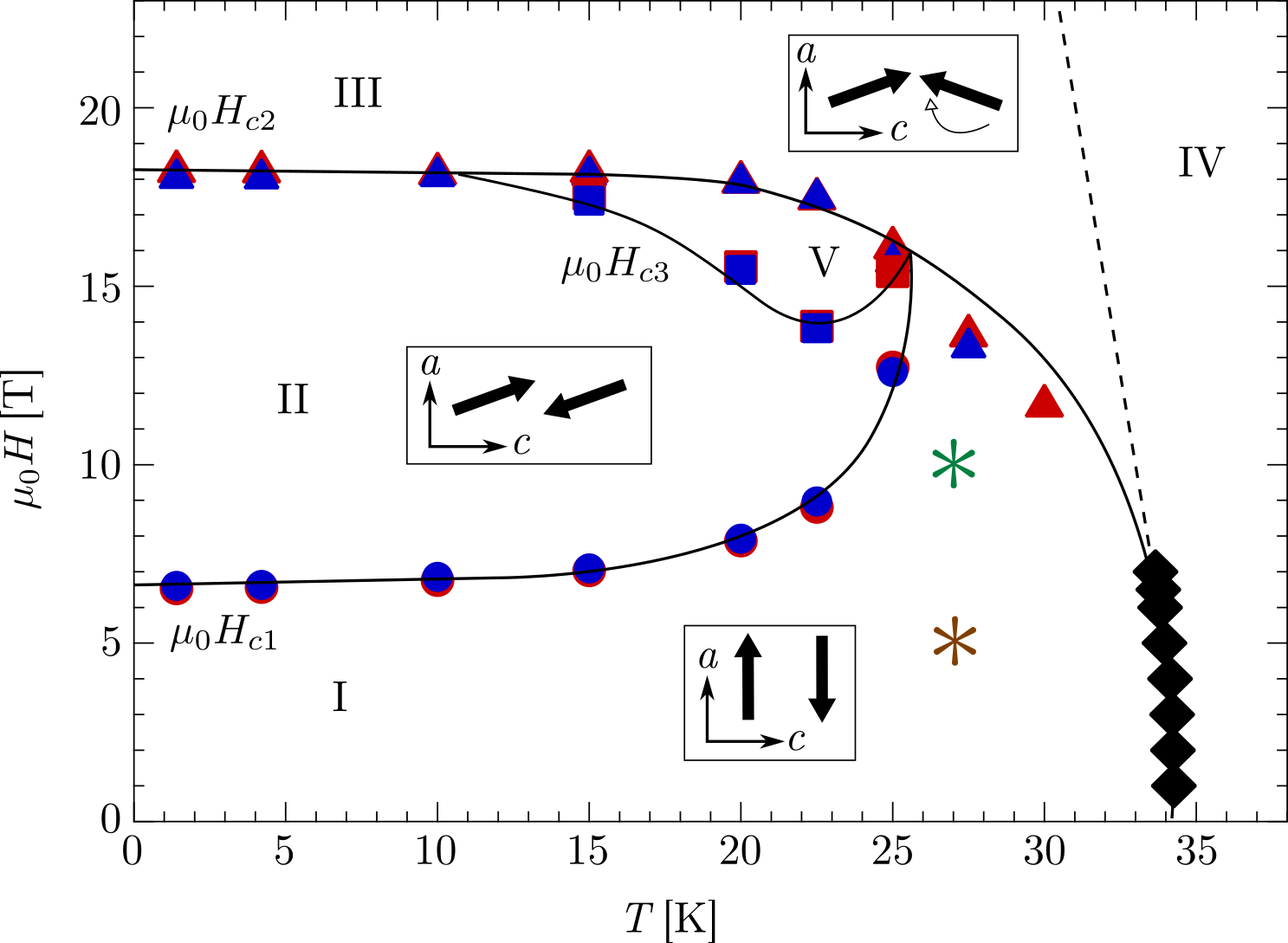}
\caption{The magnetic phase diagram of $\alpha$-Cu$_2$V$_2$O$_7$ is taken from Ref.~\onlinecite{ganatee2}.  $\mu_0 H_{c1}$ and $\mu_0 H_{c2}$ represent the spin-flop and spin-flip transitions, respectively, whereas $\mu_0 H_{c3}$ represents the transition from the spin-flop state to the intermediate state.  The green and blown asterisks denote the positions, at which the contour map in Fig.~\ref{fig4}(a) and the constant-$\mathbf{Q}$ scans in Figs.~\ref{fig4}(b) and~\ref{fig4}(c) were measured, respectively.}
\label{figS10}
\end{center}
\end{figure}

To obtain the spin dynamics of the magnons in $\alpha$-Cu$_2$V$_2$O$_7$, we follow the same procedure described in the spin-chain model. Out of the 16 magnon modes, we chose to draw the spin structure belonging to the two most intense magnon modes, i.e., the $+k$ and $-k$ modes as shown in Fig.~\ref{fig1}. In $\alpha$-Cu$_2$V$_2$O$_7$, spins along the zig-zag chain rotate clockwise for the $+k$ mode, and anticlockwise for the $-k$ mode. We note that the $\pm k$ modes refer to the shifts of the dispersion and do not imply the spin rotation. Because of the difference in the chain direction between $\alpha$-Cu$_2$V$_2$O$_7$ and the spin-chain model, the shift of the anticlockwise and clockwise modes in $\alpha$-Cu$_2$V$_2$O$_7$ is not the same as that in the spin-chain model. Figure~\ref{figS9} shows the spin dynamics of $\alpha$-Cu$_2$V$_2$O$_7$ calculated using the fitted Hamiltonian parameters at $\mathbf{q}=(0,-0.25,0)$ (left column) and $(0,0.25,0)$ (right column). When the magnetic field is applied along the $+a$ direction, the spin rotation of the $+k$ mode is in the opposite sense to the spin precession in the $bc$-plane so it requires more energy to excite this mode, whereas for the $-k$ mode the spins rotate in the same sense as the spin precession. Hence the $+k$ mode shifts upward while the $-k$ mode shifts downward consistent with the measured (Fig.~\ref{fig2}) and calculated (Fig.~\ref{figS5}) magnon dispersion.

\end{document}